\documentclass[a4paper,USenglish]{dfu}
\usepackage{graphicx}
\usepackage{listings}
\usepackage{eqnarray}
\usepackage{amsmath}
\usepackage{xspace}
\usepackage{listings}
\usepackage{enumitem}
\usepackage{color}
\usepackage{wrapfig}
\usepackage{paralist}

\usepackage[textsize=footnotesize, color=blue!20]{todonotes}

\begin{document}
%
%
%
%
%
%
%

\noindent{\sffamily\LARGE\bfseries
Data Science with Vadalog:\\ Bridging Machine Learning and Reasoning
}\\

\noindent {\large
Luigi Bellomarini$^{2,1}$, Ruslan R. Fayzrakhmanov$^1$, Georg Gottlob$^{1,3}$, Andrey Kravchenko$^1$, Eleonora Laurenza$^2$, Yavor Nenov$^1$, St\'ephane Reissfelder$^1$, Emanuel Sallinger$^1$, Evgeny Sherkhonov$^1$, Lianlong Wu$^1$
}

\medskip
\noindent {\large
$^1$ University of Oxford\\
$^2$ Banca d'Italia\\
$^3$ TU Wien
}

\bigskip
\noindent
\textbf{Abstract}. Following the recent successful examples of large technology companies, many modern
enterprises seek to build knowledge graphs to provide a unified view of
corporate knowledge and to draw deep insights using machine learning and logical reasoning. There is currently a perceived disconnect between the traditional approaches for data science, typically based on machine learning and statistical modelling, and systems for reasoning with domain knowledge. In this paper we present a state-of-the-art Knowledge Graph Management System, Vadalog, which delivers highly expressive and efficient logical reasoning and provides seamless integration with modern data science toolkits, such as the Jupyter platform. We demonstrate how to use Vadalog to perform traditional data wrangling tasks, as well as complex logical and probabilistic reasoning. We argue that this is a significant step forward towards combining machine learning and reasoning in data science.


\newcommand{\rel}[1]{\textsc{#1}}
\newcommand{\att}[1]{\ensuremath{\mathit{#1}}}

\newcommand{\LBComments}[1]{\textbf{\textcolor[rgb]{0.00,0.80,0.00}{[LB:] #1}} }

\newcommand{\evgeny}[1]{\textcolor{brown}{\small[Evgeny: #1]}}
\newcommand{\Luigi}[1]{\textcolor{blue}{\small[Luigi: #1]}}
\newcommand{\Ele}[1]{\textcolor{red}{\small[Ele: #1]}}
\newcommand{\Andrey}[1]{\textcolor{red}{\small[Andrey: #1]}}

\newcommand{\indep}{\rotatebox[origin=c]{90}{\(\models\)}}
\newcommand{\vect}[1]{\boldsymbol{#1}}
\newcommand{\vadalog}{\textsc{vadalog}\xspace}

\newcommand{\svadalog}{\textsc{soft vadalog}\xspace}

\newcommand{\NP}{\textsc{NP}\xspace}

\newcommand{\shortpaper}[1]{#1}
\newcommand{\longpaper}[1]{}

\section{Introduction} Enterprises increasingly depend on intelligent
information systems that operationalise corporate knowledge as a unified source across system boundaries. Such systems
crucially rely on insights produced by data scientists, who use advanced data
and graph analytics together with machine learning and statistical models
to create predictive actionable knowledge from suitably preprocessed
corporate data by means of data wrangling. To maintain their competitive
edge, companies need to incorporate multiple heterogeneous sources of
information, including streams of structured or unstructured data from internal
systems (e.g., Enterprise Resource Planning, Workflow Management, and Supply
Chain Management), external streams of unstructured data (e.g., news and
social media feeds, and Common Crawl\footnote{\url{http://commoncrawl.org/}}), 
publicly available and proprietary
sources of semi-structured data (e.g., DBpedia~\cite{DBLP:journals/ws/BizerLKABCH09}, 
Wikidata~\cite{DBLP:journals/cacm/VrandecicK14}, UniProt~\cite{DBLP:journals/nar/Consortium17a}, \url{data.gov.uk}), structured data extracted from web pages using web data extraction
techniques \cite{DBLP:journals/vldb/FurcheGGSS13}, as well as internal and external knowledge bases/ontologies
(e.g., ResearchCyc\footnote{\url{http://www.cyc.com/researchcyc/}}, DBpedia~\cite{DBLP:journals/ws/BizerLKABCH09}, Wikidata~\cite{DBLP:journals/cacm/VrandecicK14}, FIBO\footnote{\url{https://spec.edmcouncil.org/static/ontology/}}). The integration of such diverse information is a non-trivial
task that presents data scientists with a number of challenges including:
the extraction and handling of big data with frequently changing content and
structure; dealing with uncertainty of the extracted data; and finding ways of
unifying the information from different sources.

Following the trend of large technological companies such as Google, Amazon,
Facebook, and, LinkedIn, it is becoming common for enterprises to integrate
their internal and external sources of information into a unified
\emph{knowledge graph}. A knowledge graph typically consists of graph-structured data to allow for smooth accommodation of changes in the structure of the data, and knowledge layers,
which encode business logic used for the validation and enrichment of data and
the uncovering of critical insights from it. 
Graph-structured data may stem from data directly exposed as graphs (e.g., RDF\footnote{\url{https://www.w3.org/RDF/}} used by triple stores such as GraphDB\footnote{\url{http://graphdb.ontotext.com/}}, Property Graphs used by graph databases like neo4j\footnote{\url{https://neo4j.com/}}, and JanusGraph\footnote{\url{http://janusgraph.org/}}) or relational or semi-structured data that exhibits graph structure.
The consolidated and
enriched knowledge graph is then processed using the standard data science
toolkit for graph analytics (including languages such as Cypher\footnote{\url{https://neo4j.com/developer/cypher-query-language/}}, SPARQL\footnote{\url{https://www.w3.org/TR/rdf-sparql-query/}}, and Gremlin\footnote{\url{https://tinkerpop.apache.org/gremlin.html}}), 
statistical analysis (using the R statistical framework), and machine learning 
(using toolkits such as Weka\footnote{\url{https://www.cs.waikato.ac.nz/ml/weka/}}, scikit-learn\footnote{\url{http://scikit-learn.org/}}, and TensorFlow\footnote{\url{https://www.tensorflow.org/}}).

The creation of a coherent knowledge graph from multiple sources of
unstructured, semi-structured, and structured data is a challenging task that
requires techniques from multiple disciplines. \emph{Entity resolution}~\cite{DBLP:books/daglib/0030287} is
used to combine multiple sources of (semi-)structured data that do not share
common identifiers. The goal is to identify pairs of entities that refer to the
same real-world object and merge them into a single entity. The matching is
performed using noisy, semi-identifying information (e.g., names, addresses) and relationships, and employs specialised similarity
functions for strings, numbers, and dates, to determine the overall similarity
of two entities. \emph{Information extraction}~\cite{DBLP:journals/ftdb/Sarawagi08} 
is used for automatically
extracting structured data from unstructured sources (i.e., news and social
media feeds). Thus, for example, the news feed ``PayPal buys Hyperwallet for
\$400M'' could result into the structured statement ``acquire(PayPal,
Hyperwallet)''. Information extraction is typically combined with entity 
resolution to correctly incorporate the extracted information within an 
existing knowledge graph. 

Publicly available datasets are often equipped with ontologies which describe
relationships between entities. In such cases \emph{ontological reasoning}
needs to be applied to validate whether the results of entity resolution and 
information extraction violate any of the constraints imposed by the ontology 
as well as to enrich the data with new
information stemming from the newly produced facts. Further note that,
unsurprisingly, the use of \emph{machine learning} is pervasive throughout the
stages of the data scientist's workflow: from semantically annotating web page
elements during web data extraction, through deciding whether entities should
be matched during entity resolution, to predicting numerical trends during data
analytics over the knowledge graph. Finally, observe that although
\emph{uncertainty} is intrinsic to many of the tasks in the data scientist's
workflow, it is typically resolved by the means of a threshold. For example,
during entity resolution, the similarity of the attributes of two entities is
typically converted to a probability for the two entities to be the same, and
they are matched if the probability exceeds a certain threshold. Similarly, the
information extraction stage typically associates output facts with level of
uncertainty stemming from the extraction process, but likewise to the case of
entity resolution, the uncertainty is converted into a probability for a fact 
to hold, and a
hard decision is made on whether it should be included or not. Interestingly,
one can do better than that. One may want to impose levels of uncertainty using
business rules to better inform the decision of whether and how the knowledge graph
should be updated. One such rule, for example, could be that public companies
are much more likely to acquire private companies than vice-versa (the so
called \emph{reverse takeover}). Such rules can be produced by a domain expert
or learned from the data using rule learning \cite{BGPS17}. Furthermore, instead of
ignoring the uncertainty, after it is being used to determine whether to accept
a fact or a match, for example, one could alternatively incorporate this
uncertainty into the knowledge graph and propagate them into the further stages
of data wrangling and data analytics workflow.

To carry out the different stages of the described workflow data
scientists need to use and coordinate a number of tools, languages, and
technologies: for data access they require tools for web data extraction,
various data-base management systems, triple stores and graph databases; during
knowledge graph construction they require tools for entity resolution,
information extraction, ontological reasoning, and uncertainty management; and
during the analysis stage they require tools for graph analytic, machine
learning and statistical modelling. The coordination of all these tools
can be very challenging.

In this paper we present the Vadalog engine: a state-of-the-art Knowledge Graph
Management System (KGMS) that provides a unified framework for integrating the
various tools and technologies used by data scientists. Its language Vadalog is
an extension of the rule-based language Datalog \cite{DBLP:books/aw/AbiteboulHV95}, 
and can naturally
capture SQL (through support for the SQL operators), ontological reasoning in
OWL 2 QL\footnote{\url{https://www.w3.org/TR/owl2-profiles/}} and SPARQL 
(through the use of existential quantifiers), and graph
analytics (through non-trivial support for recursion and aggregation). The
declarative nature of the language makes the code concise, manageable, and
self-explanatory. The engine is fully extensible through its bindings to
different data sources and libraries. Data extensions provide access to
relational data stored in Postgres or MySQL, for example, or to graph data
stored in neo4j or Janus, or to web data using
OXPath~\cite{DBLP:journals/vldb/FurcheGGSS13}. Library extensions allow the
integration of state-of-the-art machine learning tools such as Weka,
scikit-learn, or TensorFlow. Additional integration with libraries for string
similarities and regular expressions allows for defining complex entity
resolution workflows. The engine also supports reasoning with probabilistic
data and probabilistic rules, which makes it ideal for handling uncertainty
stemming from the different stages of the data scientist's workflow. Finally,
the Vadalog engine seamlessly integrates with Jupyter: a well-known platform
for data analysts and scientists with a convenient interface for data
processing and visualisation.

The paper is organised as follows. Section~\ref{sec:language} provides an overview of the
core language. Section~\ref{sec:system} provides a system overview of the Vadalog engine.
Section~\ref{sec:workflow} describes the various features of the system within a typical
data scientist's workflow in Jupyter. Section
\ref{sec:ml} demonstrates the engine's integration with machine
learning on typical use cases. Finally, Section \ref{sec:probabilistic_reasoning} 
describes in more detail the support of the system for probabilistic reasoning. 

This paper includes, in abbreviated form, material from a number of previous papers on the topic~\cite{BGPS17,sofsem/BellomariniGPS18,amw/BellomariniGPS18,BellomariniSG18}. 
The Vadalog system is Oxford's contribution to VADA~\cite{sigmod/KonstantinouKAC17}, a joint project of the universities of Edinburgh, Manchester, and Oxford.
We reported first work on the overall VADA approach to data wrangling in \cite{FurcheGNS16}. In this paper, we focus on the Vadalog system at its core. Currently, our system fully implements the core language and is already in use for a number of industrial applications.

\section{Core Language}\label{sec:language}

Vadalog is a Datalog-based language. It belongs to the Datalog$^\pm$ family of languages that extends Datalog by existential quantifiers in rule heads, as well as by other features, and at the same time restricts its syntax in order to achieve decidability and data tractability; see, e.g.,~\cite{CaGK13,general,CGLMP10,CaGP12}. 
The logical core of the Vadalog language corresponds to {\em Warded Datalog$^\pm$}~\cite{ArGP14,GoPi15}, which
captures plain Datalog as well as SPARQL queries under the entailment regime for OWL 2 QL~\cite{glimmsparql}%
 and is able to perform ontological reasoning tasks. Reasoning with the logical core of Vadalog is computationally efficient.
Vadalog is obtained by extending Warded Datalog$^\pm$ with additional features of practical utility.
We now illustrate the logical core of Vadalog, more details about extensions can be found in ~\cite{BGPS17}.

\medskip
\noindent
%
The logical core of Vadalog relies on the notion of wardedness, which applies a restriction on how the ``dangerous'' variables of a set of existential rules are used. Note that existential rules are also known as tuple-generating dependencies (tgds), i.e., Datalog rules where existential quantification is allowed in the head. Intuitively, a ``dangerous'' variable is a body-variable that can be unified with a labelled null value when the chase algorithm is applied, and it is also propagated to the head of the rule. For example, given the set $\Sigma$ consisting of the rules
\[
P(x) \rightarrow \exists z \, R(x,z) \quad\textrm{and}\quad R(x,y) \rightarrow P(y),
\]
the variable $y$ in the body of the second rule is ``dangerous'' (w.r.t.~$\Sigma$) since starting, e.g., from the database $D = \{P(a)\}$, the chase will apply the first rule and generate $R(a,\nu)$, where $\nu$ is a null that acts as a witness for the existentially quantified variable $z$, and then the second rule will be applied with the variable $y$ being unified with $\nu$ that is propagated to the obtained atom $P(\nu)$.

Note that, throughout this paper, we will mix the ``logical'' notation shown above that is often used in papers, and the ``code''-like notation that is used in systems, such as the Vadalog system. The above example would be given as follows in Vadalog notation:

\medskip
\noindent
\verb|   r(X,Z) :- p(X).|\\
\verb|   p(Y) :- r(X,Y).|
\medskip

\noindent
The goal of wardedness is to tame the way null values are propagated during the construction of the chase instance by posing the following conditions:
(i) all the ``dangerous'' variables should coexist in a single body-atom $\alpha$, called the \textit{ward}; (ii)
the ward can share only ``harmless'' variables with the rest of the body, i.e., variables that are unified only with database constants during the construction of the chase.

%
{\em Warded Datalog$^\pm$} consists of all the (finite) sets of warded existential rules.
As an example of a warded set of rules, the following rules encode part of the OWL~2 direct semantics entailment regime for OWL 2 QL (see~\cite{ArGP14,GoPi15}):
\begin{eqnarray*}
\underline{{\rm Type}(x,y)}, {\rm Restriction}(y,z) &\rightarrow& \exists w \, {\rm Triple}(x,z,w)\\
\underline{{\rm Type}(x,y)}, {\rm SubClass}(y,z) &\rightarrow& {\rm Type}(x,z)\\
\underline{{\rm Triple}(x,y,z)}, {\rm Inverse}(y,w) &\rightarrow& {\rm Triple}(z,w,x)\\
\underline{{\rm Triple}(x,y,z)}, {\rm Restriction}(w,y) &\rightarrow& {\rm Type}(x,w).
\end{eqnarray*}
It is easy to verify that the above set is warded, where the underlined atoms are the wards. Indeed, a variable that occurs in an atom of the form ${\rm Restriction}(\cdot,\cdot)$, or ${\rm SubClass}(\cdot,\cdot)$, or ${\rm Inverse}(\cdot,\cdot)$, is trivially harmless. However, variables that appear in the first position of ${\rm Type}$, or in the first/third position of ${\rm Triple}$ can be dangerous. Thus, the underlined atoms are indeed acting as the wards. 

Reasoning in Warded Datalog$^\pm$ is PTIME-complete in data complexity \cite{ArGP14,GoPi15}. Although polynomial time data complexity is desirable for conventional applications, PTIME-hardness can be prohibitive for ``Big Data'' applications. One such example is towards building knowledge graphs that consider huge elections in the area of computational social choice \cite{aaai/CsarLPS17}. Yet, in fact, this is true even for linear time data complexity. This is discussed in more detail in \cite{BGPS17}.

This core language has a number of extensions to make it practical, among them data types, arithmetic, (monotonic) aggregation, bindings of predicates to external data sources, binding function symbols to external functions, and more. 

We will discuss \textit{monotonic aggregation} here. Vadalog supports aggregation (\emph{min}, \emph{max}, \emph{sum}, \emph{prod}, \emph{count}), by
means of an extension to the notion of monotonic aggregations~\cite{ShYZ15}, 
which allows adopting aggregation 
even in the presence of recursion while preserving monotonicity w.r.t.\ set containment. Such functionality is crucial for performing graph analytics, an example of which is shown in Section~\ref{sec:workflow}.

We will discuss some of these extensions throughout this paper. 
One of the extensions that are planned is more support consistency, in particular consistent query answering \cite{pods/ArenasBC99,icdt/ArmingPS16} as well as view updates \cite{pods/BunemanKT02,amw/GuagliardoPS13}.

\begin{wrapfigure}{R}{0.5\textwidth}
\centering
    \includegraphics[scale=0.3]{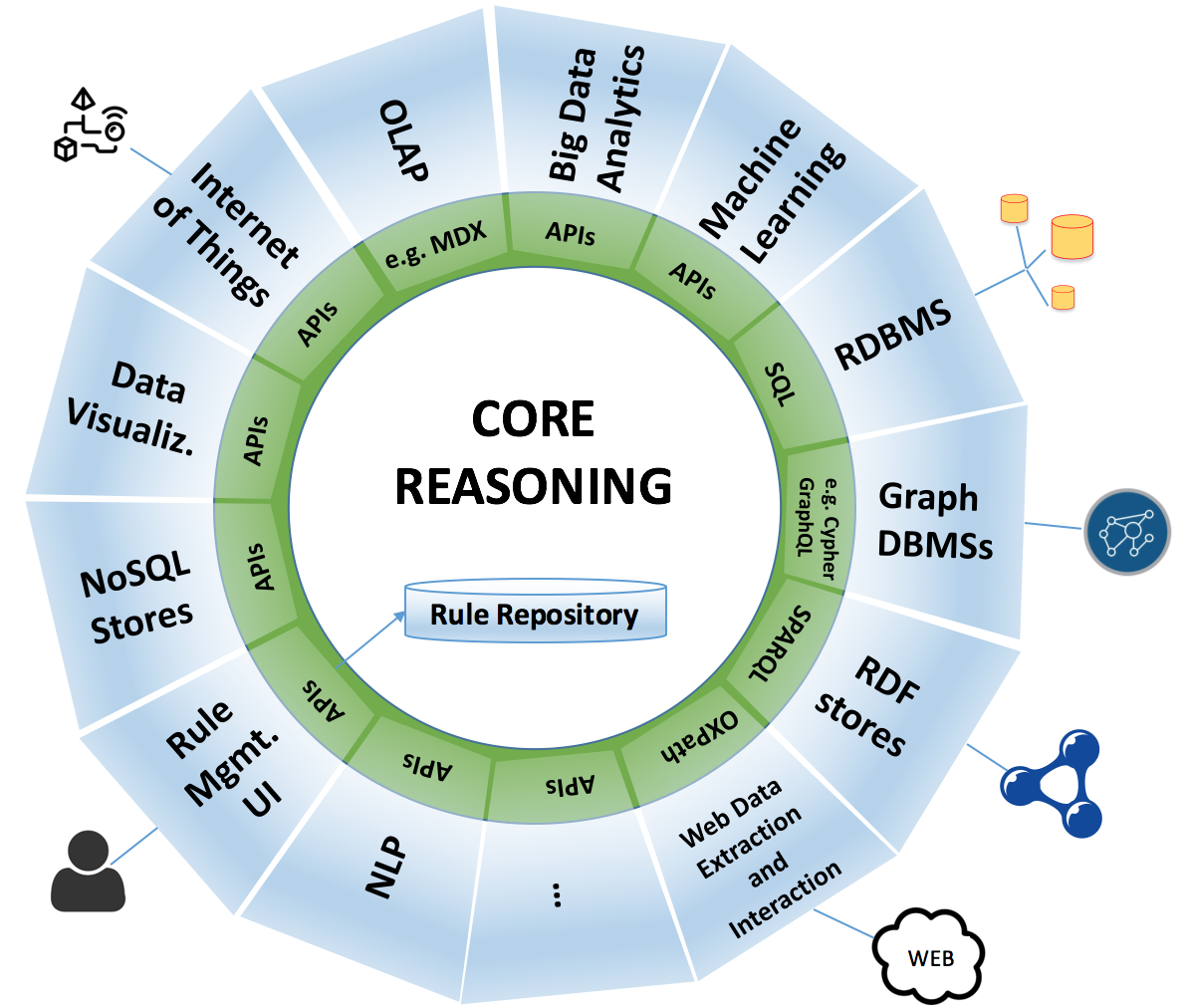} 
\vspace{-3mm}
\caption{\small{KGMS Reference Architecture~\cite{BGPS17}}}
\label{fig:architecture}
\end{wrapfigure}

\section{Core System}\label{sec:system}

The functional architecture of the Vadalog system, our KGMS, is depicted in Figure~\ref{fig:architecture}.
The knowledge graph is organised
as a repository, a collection of Vadalog rules. 
The external sources are supported by means of
\emph{transducers}, intelligent adapters that integrate the sources into the reasoning process.

The Big Data characteristics of the sources and the complex functional requirements of reasoning are tackled by leveraging the underpinnings of the core language, which are turned into practical execution strategies. In particular, 
in the reasoning algorithms devised for Warded Datalog$^\pm$, 
after a certain number of chase steps (which, in general, depends on the input database),
the chase graph~\cite{general} (a directed acyclic graph where facts are represented as nodes and the applied rules as edges)
exhibits specific periodicities and no new information, relevant to query answering, is generated.
The Vadalog system adopts an \emph{aggressive recursion and termination control} strategy,
which detects such redundancy as early as possible by combining compile-time and runtime
techniques.
In combination with a highly engineered architecture,
the Vadalog system achieves high performance and an efficient memory footprint.

At compile time, as wardedness limits the interaction between the labelled nulls, the engine
rewrites the program in such a way that joins on specific values of labelled nulls will never occur. This exploits work on schema mapping composition and optimisation \cite{pods/KolaitisPSS14,mst/KolaitisPSS18,mst/PichlerSS13,dagstuhl/Sallinger13}.

The
Vadalog
system uses a
pull stream-based approach
(or
pipeline approach), where the facts are actively requested
from the output nodes to their predecessors and so on down
to the input nodes, which eventually fetch the facts from the
data sources.   The stream approach is essential to limit the
memory consumption or at least make it predictable, so that
the system is effective for large volumes of data.
Our setting is made more challenging by the presence of
multiple  interacting  rules  in  a  single  rule  set  and  the  wide
presence of recursion. We address this by means of a specialised buffer management technique. We adopt
pervasive local
caches
in the form of wrappers to the nodes of the access plan,
where the facts produced by each node are stored.  The local
caches  work  particularly  well  in  combination  with  the  pull
stream-based approach, since facts requested by a node successor can be immediately reused by all the other successors,
without triggering further backward requests. Also, this combination  realises  an  extreme  form  of
multi-query  optimisation, where each rule exploits the facts produced by the others, whenever applicable.  To limit memory occupation, the
local caches are flushed with an
eager eviction strategy
that
detects when a fact has been consumed by all the possible
requestors and thus drops it from the memory.  Cases of actual cache overflow are managed by resorting to standard disk
swap heuristics (e.g., LRU, LFU).

More details on the Vadalog system can be found in \cite{BellomariniSG18}. The system includes many other features, such as data extraction with OXPath, which is in use with our collaborators at dblp \cite{jcdl/MichelsFLSS17}.

\section{Supporting the Data Science Workflow}\label{sec:workflow}
As the importance of data science constantly increases, the Vadalog system can support the entire spectrum of data science tasks and processes to a certain extent.
It does not however replace tools specialists like to use, but rather conveys a universal platform to integrate various approaches and tools into a unified framework.
All integrations are realised in terms of \emph{data binding primitives} and \emph{functions}.

\begin{figure}[t]
	\includegraphics[width=\textwidth]{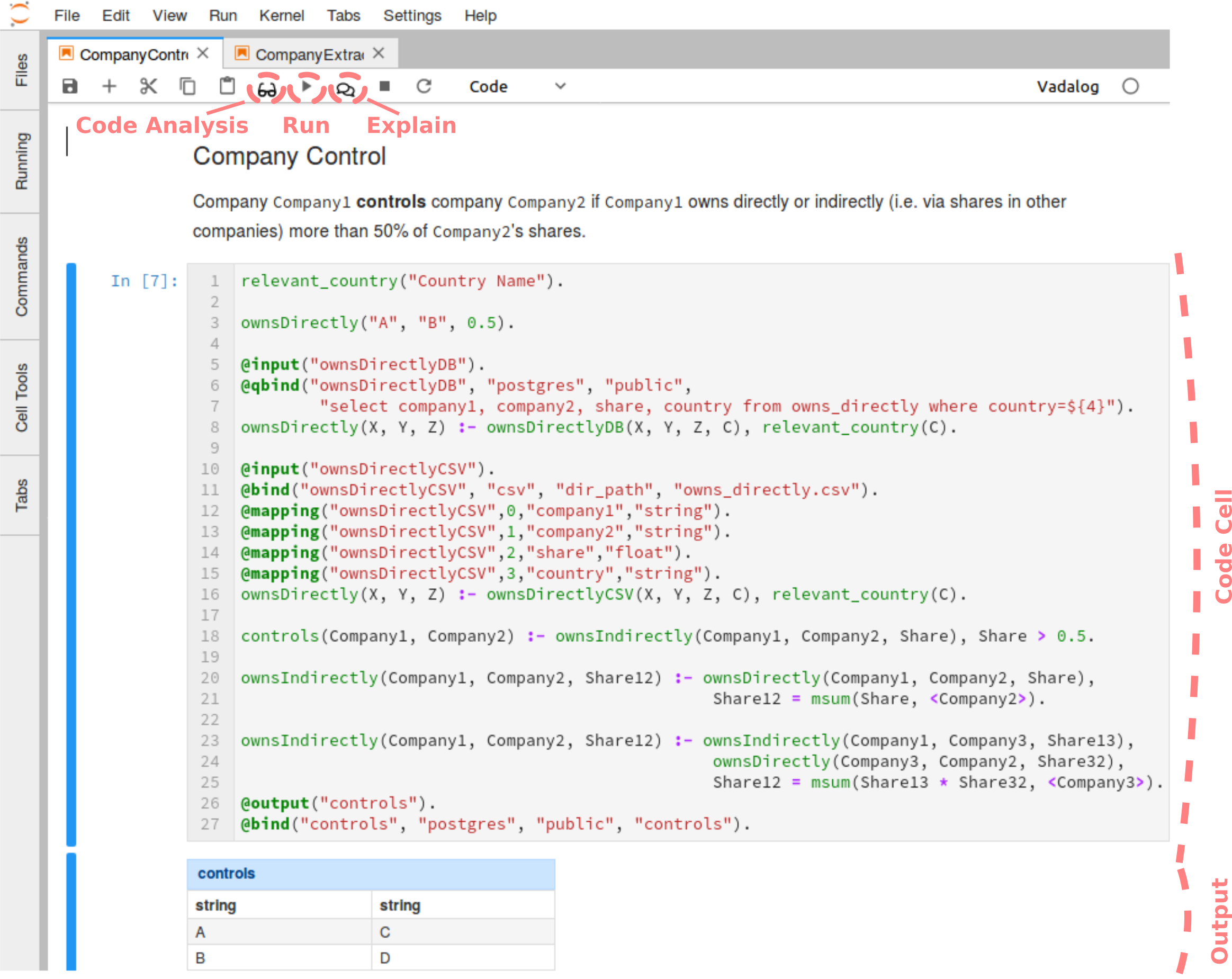}
	\caption{Example of the Vadalog program for inferring a company control indicator} \label{fig:jupyterExmpl}
\end{figure}

One such key example is the use of the UI/development platform, where Jupyter was chosen as a platform that data scientists are familiar with.
The Vadalog system has seamless integration with JupyterLab with the use of a Vadalog extension and kernel (see Figure~\ref{fig:jupyterExmpl}).
JupyterLab is a well-known platform for data analysts and scientists with a convenient interface for data processing and visualisation.
It has a multi-user support, in which dedicated resources and the environment are associated with a concrete user.
The Vadalog extension and kernel for JupyterLab give data scientists the possibility to evaluate the correctness of the program, run it, and analyse the derivation process of interesting output facts.
All output is rendered in JupyterLab's output area.

\begin{figure}[t]
	\centering
	\includegraphics[width=\textwidth]{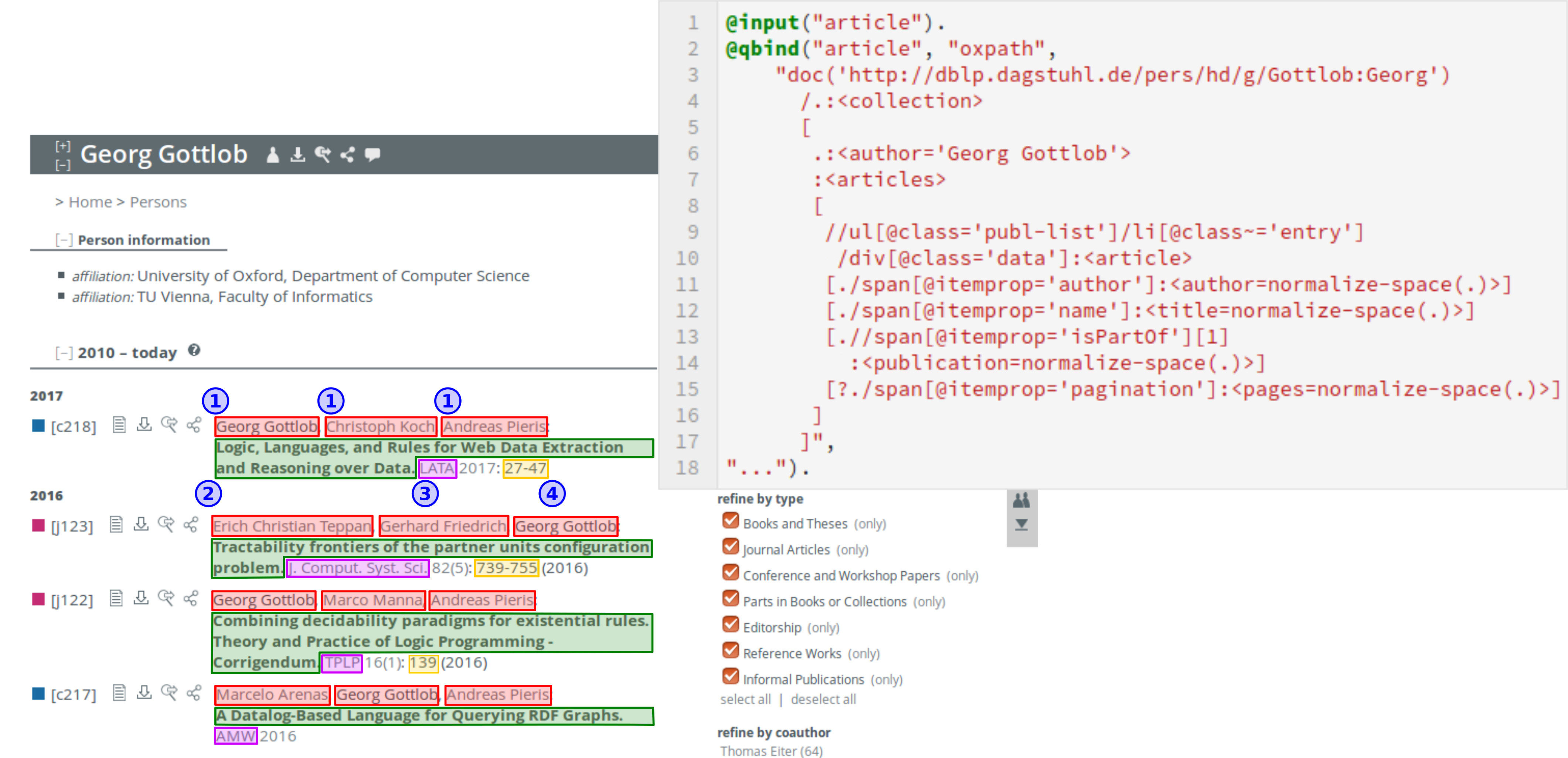}
	\caption{Integration of OXPath, a web data extraction tool}
	\label{fig:oxpath}
\end{figure}

\textbf{Data binding primitives.} Bindings give one a possibility to connect an automatic reasoning workflow with external systems for data exchange.
An external system can represent a database, framework, library or information system.
Currently Vadalog supports relational databases, such as Postgres and MySQL, and graph databases, such as neo4j.
It also has seamless integration with machine learning tools, e.g., Weka and scikit-learn (see Section~\ref{sec:mlmode1}), and a web data extraction tool, OXPath~\cite{DBLP:journals/vldb/FurcheGGSS13} (see Figure~\ref{fig:oxpath}). Other integrations are included or can be easily integrated. 
Data sources and targets can be declared by adopting \texttt{@input} and \texttt{@output} annotations.
Annotations are special facts augmenting sets of existential rules with specific behaviours.
\texttt{@input} and \texttt{@output} define the direction of facts into and from the Vadalog program, respectively.
Additional \texttt{@bind} annotation defines means for interacting with an external system.
A query bind annotation \texttt{@qbind} is a special modification of \texttt{@bind}. It supports binding predicates to queries against inputs and outputs in the external language (e.g., SQL-queries for a data source or target that supports SQL).
The first parameter of \texttt{@bind} and \texttt{@qbind} specifies a predicate the external resource is bound to;
the second parameter defines a type of the target (e.g., ``postgres'').
In case the schema of an external resource cannot be derived automatically, or should be overridden, additional \texttt{@mapping} annotation can be used to define mapping strategy for tuples between Vadalog and an external system.

In Figure~\ref{fig:jupyterExmpl}, we give a synthetic example of a Vadalog program to infer a company control indicator.
It can be formulated as follows:
\emph{A company $A$ ``controls'' company $B$ if $A$ owns directly or indirectly (i.e., via shares in other companies) more than 50\% of $B$'s shares} (lines~18--25).
As we can see, various strategies for binding external resources can be used in the Vadalog program.
For example, data tuples \texttt{ownsDirectly} can be propagated into the program from the parametric \texttt{@qbind} (lines~6-7 for Postgres via tuples \texttt{ownsDirectlyDB}) or \texttt{@bind} (line~11 for CSV via tuples \texttt{ownsDirectlyCSV}).
For \texttt{@qbind} SQL query is instantiated with the parameter from the predicate \texttt{relevant\_country} (line~8).
The query instantiation is realised within the join, in which the parameter \texttt{C} from the \texttt{relevant\_country} predicate is propagated into the fourth term of the predicate \texttt{ownsDirectlyDB}.
In contrast, in case of \texttt{@bind}, all data is streamed into the Vadalog system and filtered on-the-fly by only selecting information regarding the ``relevant country'' (line~16).
\texttt{ownsDirectly} tuples can also be specified within the program in terms of facts (line~3).
During the evaluation of the program, each derived tuple \texttt{controls} is streamed into a Postgres database as it is specified in lines~26--27.

In Figure~\ref{fig:oxpath}, we illustrate an example of binding with OXPath.
OXPath~\cite{DBLP:journals/vldb/FurcheGGSS13} is a web data extraction language, an extension of XPath 1.0 for interacting with web applications and extracting data from them.
In this example, the OXPath binding streams all articles of Georg Gottlob from dblp website into the Vadalog program.
Extracted articles can be represented as a relation \texttt{article(authors, title, publication, pages)}.
Integration with machine learning tools is discussed in the next section.

\textbf{Functions.} Besides bindings, functions provide a data scientist with a rich set of value transformations and operations for different data types supported in Vadalog.
A user can write expressions of different complexity with the use of operators and functions to perform arithmetic, manipulate strings, dates, and compare values.
Examples of supported data types are \texttt{string}, \texttt{integer}, \texttt{double}, \texttt{date}, \texttt{boolean}, \texttt{set}, and a special identifier for unknown values, \texttt{marked null}.
A data scientist can also extend the set of supported functions with those written in Python, which is enabled in the Vadalog framework.
Functions can be combined into \emph{libraries}.
For example, \texttt{@library("sim:", "simmetrics").} enables the ``simmetrics'' library in the Vadalog program, where methods can be invoked with the prefix \texttt{sim:}, as in \texttt{sim:removeDiacritics(Text)} to remove diacritics from \texttt{Text}.
We also convey libraries for building regression or classification models on-the-fly and applying those on the data derived during the automatic reasoning (see Section~\ref{sec:mlmode1}).

\begin{figure}[t]
	\centering
	\includegraphics[width=0.85\textwidth]{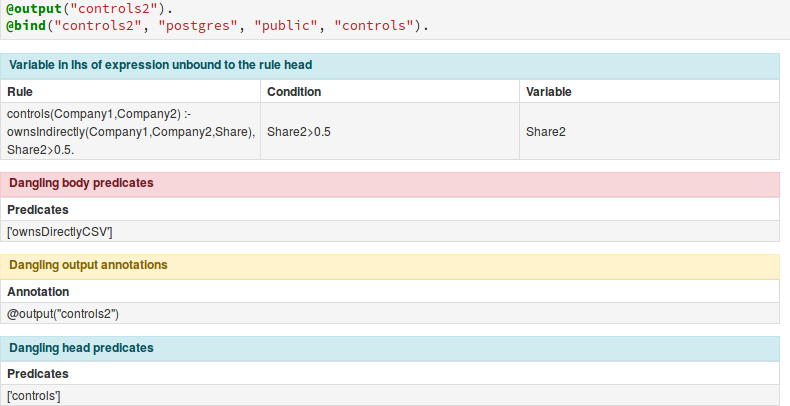}
	\caption{A screenshot depicting code analysis for an altered Vadalog program in the company control example} \label{fig:jupyterExmplAnalyse}
\end{figure}

\textbf{Code analysis.} The correctness of the program is assessed with the use of the code analysis functionality (see Figure~\ref{fig:jupyterExmplAnalyse}).
It checks whether there are essential or well-known error patterns in the program.
For example, in Figure~\ref{fig:jupyterExmplAnalyse}, we altered the original program illustrated in Figure~\ref{fig:jupyterExmpl}.
The parameter \texttt{Share} of the condition in the line~18 was replaced with \texttt{Share2}, 
lines~10--15 were commented, leaving \texttt{ownsDirectlyCSV} without the binding, and the output \texttt{controls} was changed to \texttt{controls2}.

\begin{figure}[t]
	\centering
	\includegraphics[width=\textwidth]{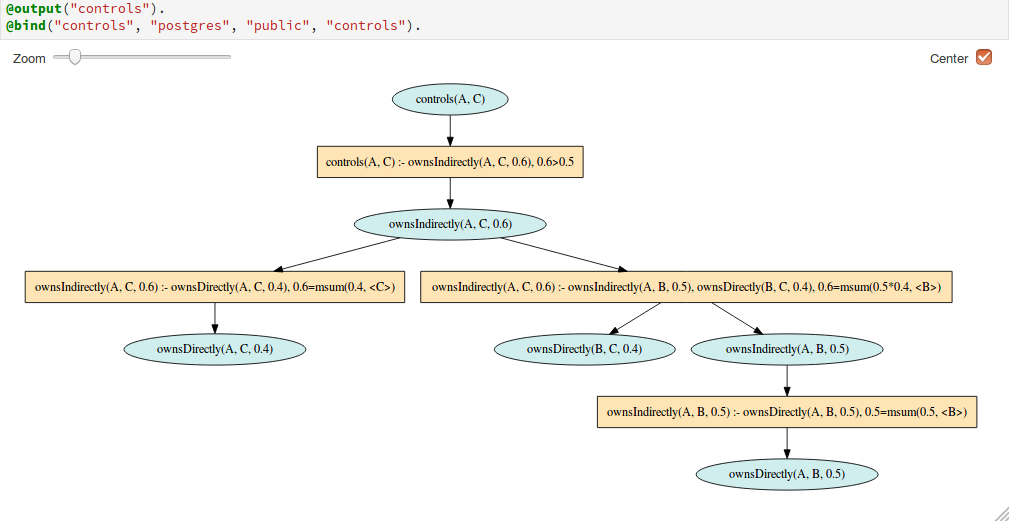}
	\caption{A screenshot of the output depicting a ``yes''-explanation for the fact \texttt{controls("A", "C")} in the company control example} \label{fig:jupyterExmplExmplain}
\end{figure}

\textbf{Fact derivation analysis.} The analysis of derivations can be performed with the use of \emph{explanations} (see Figure~\ref{fig:jupyterExmplExmplain}).
It gives an explanation of how a certain fact has been derived within the program and which rules have been triggered.

Bindings and functions make data analytics both more effective and efficient.
Vadalog directly interacts with various data sources regardless of their nature, be it a database or the Web.
Furthermore, with rich reasoning capabilities it can lift the analysis up from basic values, tuples or relations within databases to semantically rich structures, e.g., from property graphs such as of neo4j to concepts of a domain ontology. This makes the code more concise and self-explanatory.

\medskip

The Vadalog system is a universal tool which can reconcile two opposite paradigms of data scientists and domain experts, so-called ``inductive'' (or bottom-up) and ``deductive'' (or top-down) approaches.
An inductive paradigm goes along with a statement that ``patterns emerge before reasons for them become apparent''~\cite{DBLP:journals/cacm/Dhar13}.
It certainly refers to data mining and machine learning approaches which are used for deriving new knowledge and relations from data.
As all data scientists face in practice, ``all models are wrong and some are useful''~\cite[p.~208]{Box2005}, which explains problems of finding the best model given a dataset.
Furthermore, limitations related to labour intensive labelling for some machine learning algorithms can also cause incorrect or incomplete results.
Thus, knowledge of a domain expert with a deductive approach is important to correct potential errors propagated from generated models.

\section{Integrating Machine Learning}
\label{sec:ml}

\begin{wrapfigure}{R}{0.5\textwidth}
\centering
    \includegraphics[scale=0.25]{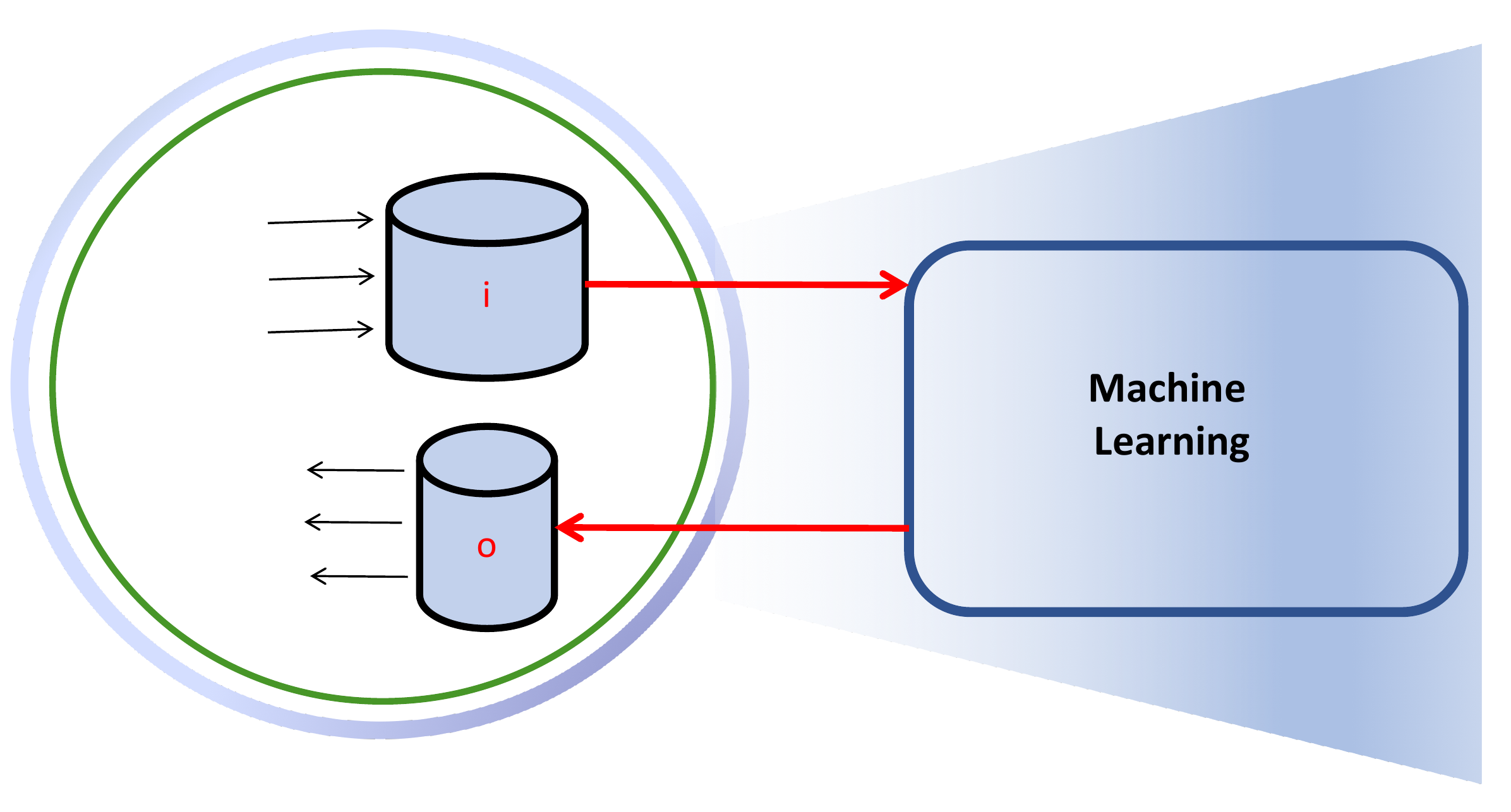} 
\vspace{-3mm}
\caption{Schematic view of the interaction between machine learning and reasoning}
\label{fig:ml-reasoning}
\end{wrapfigure}

In this section, we will discuss how to integrate machine learning directly.
We will focus on one of the approaches to machine learning integration, schematically illustrated in Figure~\ref{fig:ml-reasoning}.
In the first subsection, we will concretely talk about Weka and scikit-learn integration. The system's TensorFlow integration is similar in style to the scikit-learn integration. This will be followed in Subsection~\ref{sec:feateng} by a case study on feature engineering. We will conclude in Subsection~\ref{sec:selfml} on how to include custom ML algorithms directly into the system. 

\subsection{Direct Integration}
\label{sec:mlmode1}

\begin{figure}[t]
	\centering
	\includegraphics[width=0.8\textwidth]{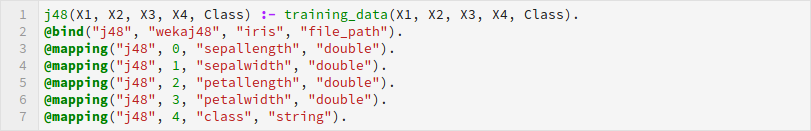}
	\caption{A snippet of Vadalog code, which demonstrates training a J48 Weka model} \label{fig:J48Train}
\end{figure}

\begin{figure}[t]
	\centering
	\includegraphics[width=0.8\textwidth]{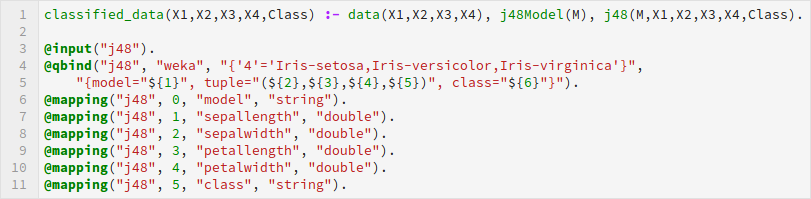}
	\caption{A snippet of Vadalog code, which demonstrates the classification phase with a trained J48 Weka model} \label{fig:J48Apply}
\end{figure}

\textbf{Weka}.
Integration with a machine learning framework, Weka, is demonstrated in Figure~\ref{fig:J48Train} and Figure~\ref{fig:J48Apply}.
Figure~\ref{fig:J48Train} illustrates the J48 model generation example for the Iris dataset.
Training data is propagated to the bound decision tree classifier associated with the predicate \texttt{j48}.
Mapping annotations specify attributes and the class of tuples streamed into the underlying machine learning algorithm.
Figure~\ref{fig:J48Apply} depicts an example of the classification process given a model \texttt{M}.
Attributes of the tuple \texttt{data} to be classified and the generated model are streamed into the underlying Weka framework via the predicate \texttt{j48}.
The results of the classification are instantiated in a relation \texttt{classified\_data}.
In the \texttt{@qbind} expression, the third parameter defines nominal attributes, a class in our case, which had index 4 in the training phase.
The fourth parameter of \texttt{@qbind} defines parameter propagation template from the predicate \texttt{j48} into the underlying model.

\medskip
\noindent
\textbf{SciPy Toolkits Machine Learning}. An external Python library such as scikit-learn can be utilised for machine learning tasks over predicates, through Vadalog Library framework. One basic linear regression example is shown below. The input consists of predicates in the form of $training\_set(ID, X, Y)$. The $sk{:}fit$ function feeds input data one by one and returns current training set size. Once sufficient training set size is reached, $sk{:}train$ function is called with a boolean return value. The last rule takes \texttt{predict} inputs one by one and retrieves output from a trained model. $\#T$ stands for boolean value true.

\medskip
{\small
 \noindent
\noindent \(@\texttt{library("sk:", "sklearn")}.\)\\
\noindent \(\texttt{training\_set("ID1", [1, 1], 2).} \)\\
\noindent \(\texttt{training\_set("ID1", [2, 2], 4).} \)\\
\noindent \(\texttt{training\_set("ID1", [3, 3], 6).} \)\\
\noindent \(\texttt{predict("ID1", [17, 17]).} \)\\
\\
\noindent \(\texttt{training\_size(ID, C) :- training\_set(ID, In, Out), C=sk:fit(ID, In, Out).} \)\\
\noindent \(\texttt{classified(ID, R) :- training\_size(ID, C), C>=3, R = sk:train(ID).} \)\\
\noindent \(\texttt{result(ID, In, Out) :-  } \)\\
\indent \(\texttt{predict(ID, In), classified(ID, \#T), Out = sk:predict(ID, In). } \)\\
}
\smallskip 
\noindent

\subsection{Case Study: Feature Engineering}
\label{sec:feateng}

We consider a case study of implementing a supervised machine learning framework and post-classification reasoning with Vadalog. Our implementation consists of three phases: \begin{inparaenum}[\bfseries(i)] \item feature extraction with Vadalog, \item interaction between Vadalog and a serialised classifier, \item post-classification reasoning. \end{inparaenum} We assume that the classifier has already been trained and serialised and for the reasons of brevity omit the description of representing a training corpus and training the classifier with Vadalog, as it can be done through a simple extension of the framework. The schematic view of the framework we implement in this case study is given in Figure~\ref{fig:ml-reasoning}.

\paragraph{Feature extraction with Vadalog.}
Consider the problem of identifying semantic blocks on a web page, such as pagination bars, navigation menus, headers, footers, and sidebars \cite{Furche2012Beryl,BerylMatwep18}. The page is represented by the DOM tree and CSS model. We represent all information contained both in the DOM and CSS as DOM facts, which are Vadalog edb predicates. An example of three DOM facts representing the \begin{inparaenum}[\bfseries(i)] \item font size of a DOM tree element with ID 100, \item its background colour, \item and the coordinates, width, and height of the corresponding CSS box \end{inparaenum} is listed below.

\medskip
{\small
 \noindent
\noindent \(\texttt{dom\_\_css\_\_fontSize("e\_100", "16px"}).\)\\
\noindent \(\texttt{dom\_\_css\_\_backgroundColor("e\_100", "rgb(229, 237, 243)"}).\)\\
\noindent \(\texttt{dom\_\_css\_\_boundingBox("e\_100", 150, 200, 450, 400, 300, 200)}.\)\\}
\smallskip

In the code snippet below we extract the feature, which computes the average font size of the sub-tree rooted at a given DOM node N, used in the navigation menu classifier, i.e., the average font size computed on a set unifying node N and all of its descendant nodes (calculated through the Start and End indices of DOM nodes).

\medskip
{\small
 \noindent
\noindent \(@\texttt{output(\texttt{"feature"})}.\)\\
\smallskip 
\noindent

\noindent \(\texttt{descendant(N,D) :-} \)\\
\indent \(\texttt{dom\_\_element(N, Start, End), dom\_\_element(D, StartD, EndD),} \)\\
\indent \(\texttt{Start < StartD, EndD < End}.\)\\

\smallskip 
\noindent

\noindent \(\texttt{feature("averageFontSize", N, FontSize) :-  dom\_\_css\_\_fontSize(N, FontSize).}\)\\

\noindent \(\texttt{feature("averageFontSize", N, FontSize) :-}\)\\
\indent \(\texttt{descendant(N,D), dom\_\_css\_\_fontSize(D, FontSize).}\)\\

\smallskip 
\noindent

\noindent \(\texttt{@post("feature", "avg(3)").}\)\\

\smallskip 
\noindent

}
\smallskip

Note that we use the \texttt{feature} namespace for the predicate, which computes this particular feature, as well as all other features used by classifiers. The feature predicates are the output of this feature extraction phase of the framework, so that they can be further passed on as input to a serialised classifier.

\paragraph{Interaction with a serialised classifier.}

All extracted features are passed on to a serialised classifier through the \texttt{@bind} operator.
For the case of web block classification, we use Weka as the machine learning library and J48 decision tree as the classifier, but the implementation of the framework in Vadalog is both library and classifier agnostic, e.g., we can seamlessly integrate Vadalog with scikit-learn, as demonstrated in Subsection~\ref{sec:mlmode1}, and the J48 decision tree classifier can also be seamlessly changed to any other classifier, e.g., an SVM. The classifications produced by the classifier are then passed back to Vadalog, also through the \texttt{@bind} operator. These classifications are in the \texttt{classification} namespace, e.g., \texttt{classification(e\_200, "navigation\_menu")} that classifies DOM node with ID 200 as a navigation menu.

\paragraph{Post-classification reasoning.}

We can now apply post-classification reasoning that cannot be easily represented by machine learning classifiers to the classifications computed in the previous phase. For example, given serialised header and footer classifiers and classifications computed in the previous phase, we can impose a constraint that a header and a footer cannot overlap.

\medskip
{\small
 \noindent

\noindent \(\texttt{header\_footer\_overlap\_constraint(N, M) :-} \)\\
\indent \(\texttt{classification(N, "header"), classification(M, "footer"),} \)\\
\indent \(\texttt{no\_overlap(N, M).} \)\\

\smallskip 
\noindent

}
\smallskip

\subsection{Direct Use of Algorithms}
\label{sec:selfml}
In case no external support is available, or users want to adapt and tie their algorithms closer to the knowledge graph, a number of Machine Learning algorithms can be directly implemented in Vadalog. Note that this is a complementary alternative -- in case algorithms should be used out-of-the-box based on existing systems and approaches, and no modification or close interaction with the knowledge graph is required, it is certainly a good idea to use such external systems and algorithms as described in Section~\ref{sec:mlmode1}. Taking advantage of the declarative programming paradigm, it requires only concisely expressing the logic of the definition, instead of explicitly describing the exact algorithm. As a result, the program is easy for modification, verification or parallel execution. The application areas include but are not limited to clustering, anomaly detection, and weekly supervised learning.

We will use DBSCAN (Density-based spatial clustering of applications with noise) algorithm as a simple example \cite{ester1996density}. 
Two main parameters of DBSCAN are $eps$ (distance threshold) and $minPts$ (minimal number of points for a dense region). The input is a set of points $p(ID, X, Y)$, $ID$ is a sequential number representing an identifier.

\begin{align*}
&eps(0.11), \ minPts(5),\\
&p(1, 0.697, 0.460), \ p(2, 0.774, 0.376), \ {\ldots}\\
\end{align*}

Two points are in a \emph{neighbourhood} if their Euclidean distance is less than $eps$. The \emph{neighbourhood number} is obtained through aggregation as below.

\begin{align*}
& p(A, X_A, Y_A), p(B, X_B, Y_B), C=\sqrt{(X_A-X_B)^2+(Y_A-Y_B)^2} \\
& \rightarrow point\_pairs (A, B, C). \\
& point\_pairs(A, B, C), eps(E), C<=E \rightarrow neighbourhood(A, B).\\
& neighbourhood(A, B), J=mcount(B) \rightarrow neighbourhood\_count(A, J). \\
& neighbourhood\_count(A, J), K=max(J) \rightarrow neighbourhood\_number(A, K).\\
\end{align*}

Different types of points, i.e., core, border and noise, are defined as follows.
\label{ex:dbscan_point_types}
\begin{align*}
&neighbourhood\_number(A, K), minPts(M), K>=M \rightarrow core\_point(A).\\
&\neg core\_point(A), core\_point(B), neighbourhood(A, B) \rightarrow border\_point(A).\\
&neighbourhood\_number(A, K), \neg core\_point(A), \neg border\_point(A) \\
&\rightarrow noise\_point(A).\\
\end{align*}

Notions of \emph{density reachability} and \emph{connectivity} are defined below.
\label{ex:dbscan_reach_connect}
\begin{align*}
&core\_point(A), neighbourhood(A, B) \rightarrow directly\_reachable(A, B).\\
&directly\_reachable(A, B) \rightarrow reachable(A, B).\\
&reachable(A, C), directly\_reachable(C, B) \rightarrow reachable(A, B).\\
&reachable(C, A), reachable(C, B) \rightarrow connected(A, B).
\end{align*}

The goal of density clustering process is to find point pairs that satisfy both \emph{connectivity} and \emph{maximality properties}, respectively:
\label{ex:dbscan_clustering}
\begin{align*}
connected(A, X) \rightarrow cluster(A, X). \\
reachable(A, X) \rightarrow cluster(A, X). & 
\end{align*}

The cluster is identified by the point (from this cluster) which has the minimal ID number.
This is achieved by the post-processing instruction, \texttt{@post}, which takes the minimum value for the second term (position) of the relation $cluster$, grouping by the first term (position).

\verb|@output("cluster"). @post("cluster", "min(2)").|\\
Output Example: \verb|cluster(1,1). cluster(2,1). cluster(3,3).|

\section{Probabilistic Reasoning}
\label{sec:probabilistic_reasoning}

In the design of winning data science solutions, it is more and more clear that completely neglecting domain knowledge
and blindly relying only on inductive models (i.e., with parameters learnt from data) easily leads to sub-optimal results,
subject to overfitting when not to wrong conclusions. 
Thus, data scientists tend to integrate inductive reasoning
with deductive approaches, complementing and when it is the case overruling
machine learning models with domain knowledge.

In the Vadalog system, we introduce \emph{probabilistic knowledge graphs}, a valuable tool to craft a new kind of data science solutions
where statistical models incorporate and are driven by the description of the domain knowledge.

Combining uncertainty and logic to describe rich uncertain relational structures is not new and has been the primary focus 
of Statistical Relational Learning (SRL)~\cite{GeTa07,Raed08}. 
One prominent representative of this area is Markov Logic Networks (MLN)~\cite{RiDo06}, which allow to describe relational structures in terms of first-order logic. 
A number of algorithms for exact and approximate reasoning in MLNs and other SRL models~\cite{bach:jmlr17,FiBrRe15} have been proposed, 
and systems built such as Alchemy~\cite{RiDo06}, Tuffy~\cite{NRDS11} and SlimShot~\cite{GrSu16}.
MLNs have been successfully applied in natural language processing~\cite{PoDo10},
ontology matching~\cite{ABS12}, record linkage~\cite{SiD06}, and so on.
Yet, one common limitation of SRL models is their logical reasoning side: logic in SRL is not utilised for deducing new knowledge,
 but rather serves the role of a constraint language. Systems that can be built on top of these models
 are hence of very limited applicability in data science tasks.
 
Consider the following example.

\begin{example}
\label{ex:first}
Let  $G$ be a knowledge graph, which contains the following facts about the ownership and link relationships between companies, 
augmented with a Vadalog program composed of rules (1) and (2):
\begin{align*}
&{\rm Own}(a,b,0.4), {\rm Own}(b,c,0.5), {\rm Own}(a,d,0.6), {\rm Own}(d,c,0.5),\\
&{\rm Linked}(a,b),{\rm Linked}(a,d),{\rm Linked}(b,c),{\rm Linked}(d,c)\\[2mm]
(1) \ & {\rm Own}(x,y,s), s>0.2 \rightarrow {\rm Linked}(x,y), \\
(2) \ & 0.8 ::  {\rm Own}(x,y,s), {\rm Own}(y,z,t), w={\rm sum}(s \cdot t) \rightarrow {\rm Own}(x,z,w) .
\end{align*}
\end{example}

Rule (1) expresses that company \(x\) is linked to \(y\) if \(x\) owns directly or indirectly more than \(20\%\) of \(y\)'s shares. Rule (2) is a recursive rule with an aggregate operator and expresses indirect shareholding: 
when \(x\) owns a number of companies \(y\), each holding a different share \(t_y\) of \(z\), then \(x\) owns \(\sum_y(s \cdot t_y)\) of \(z\).
An example of a ``traditional'' logical reasoning task is answering the following question over $G$: \emph{``which companies are linked to $a$?''}. 
The result of the reasoning task is the companies $b$ and $d$, as directly specified by $G$, and, additionally, $c$, which is implied by the
program. Indeed, by Rule (2) we first derive the fact ${\rm Own}(a,c,0.5)$, as $0.4 \times 0.5 + 0.6 \times 0.5 = 0.5$, and thus, by Rule (1), we deduce ${\rm Linked}(a,c)$.

However, here we are in an uncertain setting: Rule (2) is not definitive but holds with a certain probability.
We say that $G$ is a \emph{probabilistic knowledge graph}.
Probabilistic reasoning on $G$ would then consist in answering queries over such uncertain logic programs, i.e., when we can only access a distribution of the entailed facts.
The answer to the question ---which companies are linked to $a$--- would contain companies $b$ and $d$ with probability one
and $c$ with some probability $p$ depending on the ``ownership distance'' between \(a\) and \(c\). 

In spite of its high relevance, surprisingly, none of the exiting KGMSs allow for uncertain reasoning, crucial in many contexts.
The Vadalog system aims at filling this gap.

The Vadalog system provides a form of hybrid logic-probabilistic reasoning, where logical inference
is driven and aided by statistical inference.
We adopt the novel notion of \emph{probabilistic knowledge graph},
and propose Soft Vadalog, an extension to Vadalog
with soft, weighted rules (such as the ones used in Example~\ref{ex:first}) 
for representing and supporting uncertain reasoning in the Vadalog system.
A Soft Vadalog program is a template for a reason-tailored statistical model,
namely the \emph{chase tree}, the semantics of which is based on a probabilistic
version of the \emph{chase} procedure, a family of algorithms used in databases to enforce logic rules by generating the entailed facts.

In particular, the system adopts the \emph{MCMC-chase} algorithm: a combination of a Markov chain Monte Carlo method with the chase.
The application of the chase is guided by the MCMC, so that logical and statistical inference are
performed in the same process. We will report about these achievements soon.

\medskip
\noindent
\textbf{Acknowledgements.} This work is supported by the EPSRC programme grant EP/M025268/1. The Vadalog system is IP of the University of Oxford.

\bibliographystyle{plain}
\bibliography{main,biblio_prob}

\begin{thebibliography}{10}

\bibitem{DBLP:books/aw/AbiteboulHV95}
Serge Abiteboul, Richard Hull, and Victor Vianu.
\newblock {\em Foundations of Databases}.
\newblock Addison-Wesley, 1995.

\bibitem{ABS12}
Sivan Albagli, Rachel Ben{-}Eliyahu{-}Zohary, and Solomon~Eyal Shimony.
\newblock Markov network based ontology matching.
\newblock {\em J. Comput. Syst. Sci.}, 78(1):105--118, 2012.

\bibitem{pods/ArenasBC99}
Marcelo Arenas, Leopoldo~E. Bertossi, and Jan Chomicki.
\newblock Consistent query answers in inconsistent databases.
\newblock In {\em {PODS}}, pages 68--79. {ACM} Press, 1999.

\bibitem{ArGP14}
Marcelo Arenas, Georg Gottlob, and Andreas Pieris.
\newblock Expressive languages for querying the semantic web.
\newblock In {\em PODS}, pages 14--26, 2014.

\bibitem{icdt/ArmingPS16}
Sebastian Arming, Reinhard Pichler, and Emanuel Sallinger.
\newblock Complexity of repair checking and consistent query answering.
\newblock In {\em {ICDT}}, volume~48 of {\em LIPIcs}. SD-LZI, 2016.

\bibitem{bach:jmlr17}
Stephen~H. Bach, Matthias Broecheler, Bert Huang, and Lise Getoor.
\newblock Hinge-loss {M}arkov random fields and probabilistic soft logic.
\newblock {\em Journal of Machine Learning Research (JMLR)}, 18(109):1--67,
  2017.

\bibitem{BGPS17}
Luigi Bellomarini, Georg Gottlob, Andreas Pieris, and Emanuel Sallinger.
\newblock Swift logic for big data and knowledge graphs.
\newblock In {\em {IJCAI}}, pages 2--10, 2017.

\bibitem{sofsem/BellomariniGPS18}
Luigi Bellomarini, Georg Gottlob, Andreas Pieris, and Emanuel Sallinger.
\newblock Swift logic for big data and knowledge graphs - overview of
  requirements, language, and system.
\newblock In {\em {SOFSEM}}, volume 10706 of {\em Lecture Notes in Computer
  Science}, pages 3--16. Springer, 2018.

\bibitem{amw/BellomariniGPS18}
Luigi Bellomarini, Georg Gottlob, Andreas Pieris, and Emanuel Sallinger.
\newblock The vadalog system: Swift logic for big data and enterprise knowledge
  graphs.
\newblock In {\em {AMW}}, 2018.

\bibitem{BellomariniSG18}
Luigi Bellomarini, Emanuel Sallinger, and Georg Gottlob.
\newblock The vadalog system: Datalog-based reasoning for knowledge graphs.
\newblock {\em {PVLDB}}, 11(9):975--987, 2018.

\bibitem{DBLP:journals/ws/BizerLKABCH09}
Christian Bizer, Jens Lehmann, Georgi Kobilarov, S{\"{o}}ren Auer, Christian
  Becker, Richard Cyganiak, and Sebastian Hellmann.
\newblock Dbpedia - {A} crystallization point for the web of data.
\newblock {\em J. Web Sem.}, 7(3):154--165, 2009.

\bibitem{Box2005}
George E.~P. Box, J.~Stuart Hunter, and William~G. Hunter.
\newblock {\em {Statistics for Experimenters: Design, Innovation, and
  Discovery}}.
\newblock Wiley-Interscience, 2nd ed. edition, 2005.

\bibitem{pods/BunemanKT02}
Peter Buneman, Sanjeev Khanna, and Wang~Chiew Tan.
\newblock On propagation of deletions and annotations through views.
\newblock In {\em {PODS}}, pages 150--158. {ACM}, 2002.

\bibitem{CaGK13}
Andrea Cal\`{\i}, Georg Gottlob, and Michael Kifer.
\newblock Taming the infinite chase: Query answering under expressive
  relational constraints.
\newblock {\em J. Artif. Intell. Res.}, 48:115--174, 2013.

\bibitem{general}
Andrea Cal{\`\i}, Georg Gottlob, and Thomas Lukasiewicz.
\newblock A general datalog-based framework for tractable query answering over
  ontologies.
\newblock {\em J. Web Sem.}, 14:57--83, 2012.

\bibitem{CGLMP10}
Andrea Cal{\`{\i}}, Georg Gottlob, Thomas Lukasiewicz, Bruno Marnette, and
  Andreas Pieris.
\newblock Datalog+/-: {A} family of logical knowledge representation and query
  languages for new applications.
\newblock In {\em LICS}, pages 228--242, 2010.

\bibitem{CaGP12}
Andrea Cal\`{\i}, Georg Gottlob, and Andreas Pieris.
\newblock Towards more expressive ontology languages: {T}he query answering
  problem.
\newblock {\em Artif. Intell.}, 193:87--128, 2012.

\bibitem{DBLP:books/daglib/0030287}
Peter Christen.
\newblock {\em Data Matching - Concepts and Techniques for Record Linkage,
  Entity Resolution, and Duplicate Detection}.
\newblock Data-Centric Systems and Applications. Springer, 2012.

\bibitem{DBLP:journals/nar/Consortium17a}
The~UniProt Consortium.
\newblock Uniprot: the universal protein knowledgebase.
\newblock {\em Nucleic Acids Research}, 45(Database-Issue):D158--D169, 2017.

\bibitem{aaai/CsarLPS17}
Theresa Csar, Martin Lackner, Reinhard Pichler, and Emanuel Sallinger.
\newblock Winner determination in huge elections with mapreduce.
\newblock In {\em {AAAI}}, pages 451--458. {AAAI} Press, 2017.

\bibitem{DBLP:journals/cacm/Dhar13}
Vasant Dhar.
\newblock Data science and prediction.
\newblock {\em Commun. {ACM}}, 56(12):64--73, 2013.

\bibitem{ester1996density}
Martin Ester, Hans-Peter Kriegel, J{\"o}rg Sander, Xiaowei Xu, et~al.
\newblock A density-based algorithm for discovering clusters in large spatial
  databases with noise.
\newblock In {\em Kdd}, volume~96, pages 226--231, 1996.

\bibitem{FiBrRe15}
Daan Fierens, Guy~Van den Broeck, Joris Renkens, Dimitar~Sht. Shterionov, Bernd
  Gutmann, Ingo Thon, Gerda Janssens, and Luc~De Raedt.
\newblock Inference and learning in probabilistic logic programs using weighted
  boolean formulas.
\newblock {\em {TPLP}}, 15(3):358--401, 2015.

\bibitem{DBLP:journals/vldb/FurcheGGSS13}
Tim Furche, Georg Gottlob, Giovanni Grasso, Christian Schallhart, and
  Andrew~Jon Sellers.
\newblock {OXPath}: A language for scalable data extraction, automation, and
  crawling on the deep web.
\newblock {\em VLDB Journal}, 22(1):47--72, 2013.

\bibitem{FurcheGNS16}
Tim Furche, Georg Gottlob, Bernd Neumayr, and Emanuel Sallinger.
\newblock Data wrangling for big data: Towards a lingua franca for data
  wrangling.
\newblock In {\em {AMW}}, 2016.

\bibitem{Furche2012Beryl}
Tim Furche, Giovanni Grasso, Andrey Kravchenko, and Christian Schallhart.
\newblock Turn the page: Automated traversal of paginated websites.
\newblock In {\em ICWE}, pages 332--346, 2012.

\bibitem{GeTa07}
Lise Getoor and Ben Taskar.
\newblock {\em Introduction to Statistical Relational Learning (Adaptive
  Computation and Machine Learning)}.
\newblock The MIT Press, 2007.

\bibitem{glimmsparql}
Birte Glimm, Chimezie Ogbuji, S~Hawke, I~Herman, B~Parsia, A~Polleres, and
  A~Seaborne.
\newblock {SPARQL} 1.1 entailment regimes. {W3C} {Recommendation} 21 {March}
  2013, 2013.

\bibitem{GoPi15}
Georg Gottlob and Andreas Pieris.
\newblock Beyond {SPARQL} under {OWL} 2 {QL} entailment regime: Rules to the
  rescue.
\newblock In {\em IJCAI}, pages 2999--3007, 2015.

\bibitem{GrSu16}
Eric Gribkoff and Dan Suciu.
\newblock Slimshot: In-database probabilistic inference for knowledge bases.
\newblock {\em {PVLDB}}, 9(7):552--563, 2016.

\bibitem{amw/GuagliardoPS13}
Paolo Guagliardo, Reinhard Pichler, and Emanuel Sallinger.
\newblock Enhancing the updatability of projective views.
\newblock In {\em {AMW}}, volume 1087 of {\em {CEUR} Workshop Proceedings}.
  CEUR-WS.org, 2013.

\bibitem{pods/KolaitisPSS14}
Phokion~G. Kolaitis, Reinhard Pichler, Emanuel Sallinger, and Vadim Savenkov.
\newblock Nested dependencies: structure and reasoning.
\newblock In {\em {PODS}}, pages 176--187. {ACM}, 2014.

\bibitem{mst/KolaitisPSS18}
Phokion~G. Kolaitis, Reinhard Pichler, Emanuel Sallinger, and Vadim Savenkov.
\newblock Limits of schema mappings.
\newblock {\em Theory Comput. Syst.}, 62(4):899--940, 2018.

\bibitem{sigmod/KonstantinouKAC17}
Nikolaos Konstantinou, Martin Koehler, Edward Abel, Cristina Civili, Bernd
  Neumayr, Emanuel Sallinger, Alvaro A.~A. Fernandes, Georg Gottlob, John~A.
  Keane, Leonid Libkin, and Norman~W. Paton.
\newblock The {VADA} architecture for cost-effective data wrangling.
\newblock In {\em {SIGMOD}}. {ACM}, 2017.

\bibitem{BerylMatwep18}
Andrey Kravchenko, Ruslan~R. Fayzrakhmanov, and Emanuel Sallinger.
\newblock Web page representations and data extraction with {BERyL} (in press).
\newblock In {\em Proceedings of {MATWEP'18}}, page~8. {Springer}, 2018.

\bibitem{jcdl/MichelsFLSS17}
Christopher Michels, Ruslan~R. Fayzrakhmanov, Michael Ley, Emanuel Sallinger,
  and Ralf Schenkel.
\newblock Oxpath-based data acquisition for dblp.
\newblock In {\em {JCDL}}, pages 319--320. {IEEE} CS, 2017.

\bibitem{NRDS11}
Feng Niu, Christopher R{\'{e}}, AnHai Doan, and Jude~W. Shavlik.
\newblock Tuffy: Scaling up statistical inference in markov logic networks
  using an {RDBMS}.
\newblock {\em {PVLDB}}, 4(6):373--384, 2011.

\bibitem{mst/PichlerSS13}
Reinhard Pichler, Emanuel Sallinger, and Vadim Savenkov.
\newblock Relaxed notions of schema mapping equivalence revisited.
\newblock {\em Theory Comput. Syst.}, 52(3):483--541, 2013.

\bibitem{PoDo10}
Hoifung Poon and Pedro~M. Domingos.
\newblock Unsupervised ontology induction from text.
\newblock In {\em {ACL}}, pages 296--305, 2010.

\bibitem{Raed08}
Luc~De Raedt.
\newblock {\em Logical and Relational Learning: From ILP to MRDM (Cognitive
  Technologies)}.
\newblock Springer-Verlag, Berlin, Heidelberg, 2008.

\bibitem{RiDo06}
Matthew Richardson and Pedro~M. Domingos.
\newblock Markov logic networks.
\newblock {\em Machine Learning}, 62(1-2):107--136, 2006.

\bibitem{dagstuhl/Sallinger13}
Emanuel Sallinger.
\newblock Reasoning about schema mappings.
\newblock In {\em Data Exchange, Information, and Streams}, volume~5 of {\em
  Dagstuhl Follow-Ups}, pages 97--127. SD-LZI, 2013.

\bibitem{DBLP:journals/ftdb/Sarawagi08}
Sunita Sarawagi.
\newblock Information extraction.
\newblock {\em Foundations and Trends in Databases}, 1(3):261--377, 2008.

\bibitem{ShYZ15}
Alexander Shkapsky, Mohan Yang, and Carlo Zaniolo.
\newblock Optimizing recursive queries with monotonic aggregates in deals.
\newblock In {\em ICDE}, pages 867--878, 2015.

\bibitem{SiD06}
Parag Singla and Pedro~M. Domingos.
\newblock Entity resolution with markov logic.
\newblock In {\em {ICDM}}, pages 572--582, 2006.

\bibitem{DBLP:journals/cacm/VrandecicK14}
Denny Vrandecic and Markus Kr{\"{o}}tzsch.
\newblock Wikidata: a free collaborative knowledgebase.
\newblock {\em Commun. {ACM}}, 57(10):78--85, 2014.

\end{thebibliography}

\end{document}